\documentclass[onecolumn,useAMS,usenatbib]{mnras}
\usepackage{graphicx}
\usepackage{subfigure}
\usepackage{xcolor}
\usepackage{mathtext,bbm,amsmath,amsfonts,amssymb,indentfirst,syntonly,graphicx}
\usepackage{mathtools}
\usepackage{slashbox}
\usepackage[english]{babel}
\usepackage{calc}
\usepackage{tikz}
\usepackage[T1]{fontenc}
\usepackage{ae,aecompl}
\usepackage{latexsym,graphicx,bbm}
\usepackage{enumerate}
\usepackage{longtable}
\usepackage{supertabular}
\usepackage{multirow}

\title[probing the baryon mass fraction in IGM with FRBs]{Probing the baryon mass fraction in IGM and its redshift evolution with fast radio bursts using Bayesian inference method}
\author[H.-N. Lin and R. Zou]
{Hai-Nan Lin$^{1,2}$, Rui Zou$^{1,2}$\thanks{Corresponding author: zourui@stu.cqu.edu.cn}\\
$^{1}$Department of Physics, Chongqing University, Chongqing 401331, China\\
$^{2}$Chongqing Key Laboratory for Strongly Coupled Physics, Chongqing University, Chongqing 401331, China\\}
\begin{document}

\date{Accepted 2023; Received 2023; in original form 2023}

\pagerange{\pageref{firstpage}--\pageref{lastpage}} \pubyear{2023}

\maketitle

\label{firstpage}

\begin{abstract}
  We investigate the fraction of baryon mass in intergalactic medium ($f_\mathrm{IGM}$), using 18 well-localized FRBs in the redshift range $z\in (0.0039,0.66)$. We construct a five-parameter Bayesian inference model, with the probability distributions of dispersion measures (DM) of IGM and host galaxy properly taken into account. To check the possible redshift evolution, we parameterize $f_\mathrm{IGM}$ as a mildly evolving function of redshift, $f_\mathrm{IGM}=f_\mathrm{IGM,0}[1+\alpha z/(1+z)]$. By simultaneously constraining five parameters, we get $f_\mathrm{IGM,0} = 0.92^{+0.06}_{-0.12}$ and $\alpha = 0.49^{+0.59}_{-0.47}$, and the median value of DM of host galaxy is $\exp(\mu)=72.49^{+33.31}_{-25.62}~{\rm pc ~ cm ^ {-3}}$. By fixing two parameters which can be constrained independently with other observations, we obtain $\alpha =0.11^{+0.24}_{-0.27}$ in the three-parameter fit, which is consistent with zero within $1\sigma$ uncertainty. Monte Carlo simulations show that even 300 FRBs are not enough to tightly constrain five parameters simultaneously. This is mainly caused by the correlation between parameters. Only if two parameters are fixed, 100 FRBs are necessary to achieve unbiased constraints on the remaining parameters.
\end{abstract}

\begin{keywords}
  fast radio bursts  --  intergalactic medium  --  cosmological parameters
\end{keywords}

\section{Introduction}\label{sec:introduction}

Fast radio bursts (FRBs) are milisecond-duration radio transients that randomly happen in the sky \citep{Cordes:2019cmq,Petroff:2019tty,Xiao:2021omr,Zhang:2022uzl}. FRBs were first detected by the Parkes telescope in 2007 \citep{Lorimer:2007qn}, and several hundred FRB sources have been collected till now \citep{Petroff:2016tcr, CHIMEFRB:2021srp}. Among them only one is confirmed to originate from the Milky Way \citep{CHIMEFRB:2020abu,Bochenek:2020zxn}, while the others are expected to have extragalactic origins \citep{Keane:2016yyk,Chatterjee:2017dqg,Tendulkar:2017vuq}. Generally, FRBs can be divided into two types, i.e., repeaters and non-repeaters, distinguished by whether they flash only once or more. But it is still unclear if the apparently non-repeaters will repeat or not in the future. Since the discovery of the first repeating event FRB 121102 with a redshift measurement of $z = 0.19$ \citep{Spitler:2016dmz,Scholz:2016rpt,Chatterjee:2017dqg,Marcote:2017wan,Tendulkar:2017vuq}, FRBs have inspired scientists to study the underlying physics and retract the origin of the mysterious pulses. In recent years, an increasing number of FRBs have been detected thanks to the improvement of detection techniques and the operation of new telescopes, such as the Canadian Hydrogen Intensity Mapping Experiment \cite[CHIME,][]{Amiri_2018} and the Five-hundred-meter Aperture Spherical Telescope \cite[FAST,][]{Nan:2011um}, which gives us a new chance to deep investigate the Universe.

FRBs are very luminous and most of them have large dispersion measure (DM), indicating extragalactic or even cosmological origin \citep{Lorimer:2007qn,Thornton:2013iua,Petroff:2014taa,Petroff:2016tcr}. The direct measurement of redshift of FRB 150418 confirmed this hypothesis \citep{Keane:2016yyk}. Therefore, FRBs can be used to probe cosmological parameters such as the dark matter \citep{Munoz:2016tmg}, the baryon mass density \citep{Walters:2017afr,Macquart:2020lln}, the baryon mass fraction in intergalactic medium (IGM) \citep{Li:2019klc,Li:2020qei}, the Hubble parameter \citep{Wu:2020jmx} and  Hubble constant \citep{Li:2017mek}, the cosmic curvature\citep{Li:2017mek}, the cosmic proper distance \citep{Yu:2017beg} and the reionization history of hydrogen and helium \citep{Pagano:2021zla}, etc. Besides, FRBs can also be used to test the fundamental physics, such as constraining the rest mass of photon \citep{Wu:2016brq,Bonetti:2016cpo}, testing the Einstein's equivalence principle \citep{Wei:2015hwd,Tingay:2016tgf}, and constraining the Lorentz invariance violation \citep{Wei:2021vvn}. All these applications can be achieved from the DM-redshift relation.

One problem that hinders the applications of FRBs in cosmology is the strong degeneracy between cosmological parameters and the baryon mass fraction in IGM ($f_\mathrm{IGM}$). \citet{Fukugita:1997bi} estimated that the baryons in stars and their remnants only comprise about $17\%$ of the total baryon based on numerous observations, while the remaining $83\%$ baryons are in a diffuse state in IGM. This is the so-called ``missing baryon" problem. Since then, many numerical simulations \citep{Cen:1998hc,Cen:2006by,Ferrara:2014sda} and observations \citep{Munoz:2018mll,Fukugita:2004ee,Shull:2011aa,McQuinn:2013tmc,Hill:2016dta} have been done in hope that the missing baryon problem would be alleviated. For example, \citet{Ferrara:2014sda} estimated that of $f_\mathrm{IGM} = 0.82$ at $z\leq 0.4$ and $f_\mathrm{IGM} = 0.9$ at $z\geq 1.5$, showing moderate redshift-evolution. FRBs, as the most energetic radio transients in the Universe, are excellent tools to probe the baryons. \citet{Li:2019klc} proposed a cosmology-independent method to estimate $f_\mathrm{IGM}$ using FRBs. Using this method, \citet{Li:2020qei} obtained $ f_\mathrm{IGM} = 0.84 ^{+0.16}_{-0.22}$ from five well-localized FRBs, but no evidence of redshift-dependence was found. In a word, the $f_\mathrm{IGM}$ term is still not well constrained, especially its redshift evolution.

There are several difficulties in using FRBs as probes to study the cosmology. First, the available FRB sample is not large enough. Though hundreds of FRBs have been detected, only a tiny number of them are well-localized and have redshift measurements. Second, due to the large-scale density fluctuations, the DM contribution of IGM,  $\mathrm{DM_{IGM}}$, has large uncertainty. The true value of $\mathrm{DM_{IGM}}$ may significantly deviate from the expectation \citep{Macquart:2020lln}. Finally, the DM contribution of host galaxy, $\mathrm{DM_{host}}$, is poorly known, which may be affected by many factors such as the galaxy type, the mass of host galaxy, the inclination angle of host galaxy, the star-formation rate (SFR), etc. Hence, to model $\mathrm{DM_{host}}$ is not an easy task. Some papers in the literature treat this term as a constant, or model it as SFR-dependent \citep{Ioka:2003fr,Deng:2013aga,Li:2019klc}. However, the actual value of $\mathrm{DM_{host}}$ can vary significantly from burst to burst. Numerical simulations show that $\mathrm{DM_{host}}$ follow a certain probability distribution \citep{Macquart:2020lln,Zhang:2020mgq}. Thus, properly dealing with $\mathrm{DM_{host}}$ is important when using FRBs to constrain cosmological parameters.

In this paper, we investigate the baryon mass fraction in IGM, especially its redshift evolution using well-localized FRBs. The probability distributions of $\mathrm{DM_{IGM}}$ and $\mathrm{DM_{host}}$ are properly taken into consideration. The Bayesian inference method is applied to constrain the free parameters. The rest parts of this paper are arranged as follows: In Section \ref{sec_2}, we introduce a Bayesian framework to constrain the baryon mass fraction in IGM using FRBs. The observational data and the constraining results are given in Section \ref{sec:results}. In Section \ref{sec_3}, we perform Monte Carlo simulations to check the validity of our method. Finally, discussion and conclusions are given in Section \ref{sec_4}.

\section{METHODOLOGY}\label{sec_2}

The propagation of FRBs can be easily influenced by the intervening medium between the host galaxy and the observer on earth. Most importantly, the interaction of radio waves with cold plasma leads to the delay of arriving time, i.e., the pulse with higher frequency arrives earlier than the lower one. The time delay caused by the plasma effect is proportional to a quantity named dispersion measure (DM), which can be expressed as the integral of the electron number density along the travelling path, $\mathrm{DM_{FRB}} = \int n_e \,\mathrm{d}l/(1+z) $. Physically, the total DM of an FRB can be decomposed into four primary components \citep{Gao_2014,Deng:2013aga,Macquart:2020lln},
\begin{equation}
  \mathrm{DM_{FRB}}(z) = \mathrm{DM_{MW,ISM}} + \mathrm{DM_{MW,halo}} + \mathrm{DM_{IGM}}(z) + \frac{\mathrm{DM_{host}}}{1+z} ,
\end{equation}
where $\mathrm{DM_{MW,ISM}}$ is the contribution from the Galactic interstellar medium (ISM), $\mathrm{DM_{MW,halo}}$ is the contribution from the Galactic halo, $\mathrm{DM_{host}}$ is the contribution from host galaxy in the source frame, which is weighted by $(1+z)^{-1}$ if converted to the observer frame, and $\mathrm{DM_{IGM}}$ is the contribution from IGM.

The $\mathrm{DM_{MW,ISM}}$ term can be estimated from the Galactic electron density models, such as the NE2001 model \citep{Cordes:2002wz} and the YMW16 model \citep{Yao_2017}. These two models give consistent results at high Galactic latitude, but it is shown that the YMW16 model may overestimate $\mathrm{DM_{MW,ISM}}$ at low Galactic latitude \citep{KochOcker:2021fia}. Therefore, we apply the NE2001 model to calculate $\mathrm{DM_{MW,ISM}}$. We still have poor knowledge on the exact value of the $\mathrm{DM_{MW,halo}}$. \citet{Prochaska:2019stz} estimated that it is in the range of $50-100$ $\mathrm{pc\,cm^{-3}}$. Therefore, we follow \citet{Macquart:2020lln} and conservatively assume $\mathrm{DM_{MW,halo}} = 50\,\mathrm{pc\,cm^{-3}}$. The value of $\mathrm{DM_{host}}$ may vary significantly in different sources, which will become less important at high redshift because of the $(1+z)^{-1}$ factor suppression and the domination of the $\mathrm{DM_{IGM}}$ term. But at low-redshift this term can't be ignored. The $\mathrm{DM_{IGM}}$ term is strongly redshift-dependent, which contains the information of cosmological parameters. The probability distributions of $\mathrm{DM_{host}}$ and $\mathrm{DM_{IGM}}$ are discussed below.

Based on the standard $\Lambda\mathrm{CDM}$ model, the mean value of $\mathrm{DM_{IGM}}$ at redshift $z$ can be written as \citep{Deng:2013aga,Zhang:2020ass}
\begin{equation}\label{DM_IGM_average}
  \langle{\rm DM_{IGM}}(z)\rangle=\frac{3cH_0\Omega_bf_{\rm IGM}f_e}{8\pi Gm_p}\int_0^z\frac{1+z}{\sqrt{\Omega_m(1+z)^3+\Omega_\Lambda}}dz,
\end{equation}
where $H_0$ is the Hubble constant, $\Omega_b$ is the cosmic baryon mass density, $\Omega_m$ and $\Omega_\Lambda$ are the matter density and vacuum energy density of the Universe, respectively. The three constants $c$, $G$ and $m_p$ are the speed of light, the Newtonian gravitational constant and the mass of proton, respectively. $f_e = Y_{\rm H}X_{e,{\rm H}}(z) + \frac{1}{2}Y_{\rm He}X_{e,\rm He}(z)$ denotes the extent of ionization progress of hydrogen and helium, in which $Y_{\rm H} = 0.75$ and $Y_{\rm He} = 0.25$ are the mass fractions of hydrogen and helium, $X_{e,\rm H}$ and $X_{e,\rm He}$ are their ionization fractions, respectively. Considering both hydrogen and helium are fully ionized at $z<3$ \citep{Meiksin:2007rz,Becker:2010cu}, we take $X_{e,\rm H} = X_{e,\rm He} = 1$. $f_\mathrm{IGM}$ is the fraction of baryon mass in IGM, which may slowly increase with redshift \citep{Ferrara:2014sda}, but the accurate form is still unclear. In this paper, we follow \citet{Li:2019klc} and phenomenological parameterize it as a mildly evolving function of redshift,
\begin{equation}\label{f_IGM}
  f_\mathrm{IGM}(z) = f_\mathrm{IGM,0}\left(1+\frac{\alpha z}{1+z}\right),
\end{equation}
where $f_\mathrm{IGM,0}$ is the baryon mass fraction in IGM at $z=0$, and $\alpha$ is the evolving parameter that are expected to prefer a positive value \citep{Li:2019klc,McQuinn:2013tmc}. We work in the standard $\Lambda\mathrm{CDM}$ model with the Planck 2018 parameters, i.e., $H_0=67.4~{\rm km~s^{-1}~Mpc^{-1}}$, $\Omega_m=0.315$, $\Omega_\Lambda=0.685$ and $\Omega_{b}=0.0493$ \citep{Planck:2018vyg}.

Although the mean value of $\rm DM_{IGM}$ is given in equation (\ref{DM_IGM_average}), the actual value of this term will vary around the mean due to the large-scale fluctuations. The probability density function of $\rm DM_{IGM}$ can be derived from theoretical analyses of the IGM and galaxy haloes \citep{McQuinn:2013tmc,Prochaska:2019stz}, which can be fitted using the function \citep{Macquart:2020lln,Zhang:2020xoc}
\begin{equation}\label{p_delta}
  p_{\rm IGM}(\Delta)=A\Delta^{-\beta}\exp\left[-\frac{(\Delta^{-\alpha}-C_0)^2}{2\alpha^2\sigma_{\rm IGM}^2}\right], ~~~\Delta>0,
\end{equation}
where $\Delta\equiv{\rm DM_{IGM}}/\langle{\rm DM_{IGM}}\rangle$, and $\sigma _\mathrm{IGM} = Fz^{-0.5}$ represents the effective standard deviation, with $F$ being the baryon feedback coefficient. $\alpha$ and $\beta$ are two parameters related to the inner density profile of gas in haloes, which are chosen to be $\alpha = \beta = 3$ \citep{Macquart:2020lln}. $A$ is the normalization constant, and $C_0$ is calculated to make sure that the mean of the distribution is unity.

The distribution of $\mathrm{DM_{host}}$ has limited theoretical motivation because of the absence of information about the local environment of FRB sources. It may range from several ten to several hundred $\mathrm{pc ~ cm ^ {-3}}$. For example, \citet{Xu:2021qdn} estimated the $\mathrm{DM_{host}}$ of FRB20201124A to be in the range of $10 \sim 310 ~\mathrm{pc ~ cm ^ {-3}}$, while \citet{Niu:2021bnl} estimated that of FRB20190520B to be as large as $900~{\rm pc~cm^{-3}}$. To account for the possible existence of large $\mathrm{DM_{host}}$ value, we assume that it follows the log-normal distribution \citep{Macquart:2020lln,Zhang:2020mgq},
\begin{equation}\label{p_dm_host}
  p_{\rm host}({\rm DM_{host}}|\mu,\sigma_{\rm host})=\frac{1}{\sqrt{2\pi}{\rm DM_{host}}\sigma_{\rm host}} \exp\left[-\frac{(\ln {\rm DM_{host}}-\mu)^2}{2\sigma_{\rm host}^2}\right],
\end{equation}
where $\mu$ and $\sigma_{\rm host}$ denote the mean and standard deviation of $\ln {\rm DM_{host}}$, respectively. Generally, $\mu$ and $\sigma_{\rm host}$ should be redshift dependent, but for non-repeating bursts they do not vary significantly with redshift \citep{Zhang:2020mgq}. For simplicity, we follow \citet{Macquart:2020lln} and treat the two parameters as constants.

Since the DM contributions from the Milky Way (including ${\rm DM_{MW,ISM}}$ and ${\rm DM_{MW,halo}}$) can be estimated directly, we can subtract them from the total DM. For simplicity, we define the extragalactic $\mathrm{DM}$ as
\begin{equation}\label{DM_E}
  {\rm DM_E}\equiv {\rm DM_{FRB}}-{\rm DM_{MW,ISM}}-{\rm DM_{MW,halo}}={\rm DM_{IGM}}+\frac{{\rm DM_{host}}}{1+z}.
\end{equation}
Based on the probability distributions of $\mathrm{DM_{IGM}}$ and $\mathrm{DM_{host}}$, the probability distribution of $\mathrm{DM_E}$ can be calculated as
\begin{equation}\label{eq:P_E}
  p_{\rm E}({\rm DM_E}|z)=\int_0^{(1+z)\rm DM_E}p_{\rm host}({\rm DM_{host}}|\mu,\sigma_{\rm host})p_{\rm IGM}({\rm DM_E}-\frac{\rm DM_{host}}{1+z}|F,f_\mathrm{IGM,0},\alpha)d{\rm DM_{host}}.
\end{equation}
Then the joint likelihood function of a sample of FRBs can be written as
\begin{equation}
  \mathcal{L}=\prod_{i=1}^Np_{\rm E}({\rm DM_{E,\it i}}|z_i),
\end{equation}
where $N$ is the total number of FRBs. According to the Bayesian theorem, the posterior probability density function of the free parameters can be written as
\begin{equation}
  P(F,f_\mathrm{IGM,0},\alpha,\mu,\sigma_{\rm host}|{\rm FRBs})\propto\mathcal{L}({\rm FRBs}|F,f_\mathrm{IGM,0},\alpha,\mu,\sigma_{\rm host})P_0(F,f_\mathrm{IGM,0},\alpha,\mu,\sigma_{\rm host}),
\end{equation}
where $P_0$ is the prior of the free parameters. There are five free parameters in total: three parameters related to $\mathrm {DM_{IGM}}$ ($F$, $f_\mathrm{IGM,0}$ and $\alpha$), and two parameters related to $\mathrm {DM_{host}}$  ($\mu$ and $\sigma_{\mathrm {host}}$), which will be constrained simultaneously using well-localized FRBs in the next section.

\section{Data and Results}\label{sec:results}

Till now, there are 21 well-localized extragalactic FRBs that have accurate identification of host galaxy and direct measurement of redshifts\footnote{There are 19 well-localized FRBs compiled in the FRB Host Database, http://frbhosts.org/. Two other FRBs, FRB20171020A \citep{Li:2019kim} and FRB20190520B \citep{Niu:2021bnl}, are not included in the database.}. Among them, FRB20200120E, FRB20190614D and FRB20190520B are ingored. The source of FRB20200120E is near our Galaxy, and the peculiar velocity dominates over the Hubble flow, so it has a negative spectroscopic redshift $z = -0.001$ \citep{Bhardwaj:2021xaa,Kirsten:2021llv}. FRB20190614D doesn't have a direct measurement of spectroscopic redshift, but have a photometric redshift $z\approx 0.6$ \citep{Law:2020cnm}. The $\mathrm{DM_{host}}$ of FRB20190520B is estimated to be as large as $900~{\rm pc~cm^{-3}}$ \citep{Niu:2021bnl}, which is much larger than the normal FRBs, so this FRB is also excluded in our sample. The remaining 18 FRBs have well measured spectroscopic redshifts, and their main properties are listed in Table \ref{tab:host}, which will be used to constrain cosmological parameters.

\begin{table}
  \centering
\dag  \caption{The catalog of 18 well-localized FRBs. $\rm{DM_{MW,ISM}}$ is calculated using NE2001 model, and ${\rm DM_E}$ is calculated with equation (\ref{DM_E}) by assuming ${\rm DM_{MW,halo}} = 50~{\rm pc~cm^{-3}}$.}\label{tab:host}
  {\begin{tabular}{ccccccccl} 
  \hline\hline 
  FRBs & RA & Dec & ${\rm DM_{FRB}}$ & ${\rm DM_{MW,ISM}}$ & ${\rm DM_E}$ & $z_{\rm sp}$ & repeat? & reference\\
  & [ $^{\circ}$ ] & [ $^{\circ}$ ] & [${\rm pc~cm^{-3}}$] & [${\rm pc~cm^{-3}}$] & [${\rm pc~cm^{-3}}$] & & \\
  \hline
  20121102A & $82.99$ & $33.15$ &557.00 &157.60 &349.40 &0.1927 & Yes & \citet{Chatterjee:2017dqg}\\
  20171020A & $22.15$ & $-19.40$ &114.10 &38.00 &26.10 &0.0087 & No & \citet{Li:2019kim}\\
  20180301A & $93.23$ & $4.67$ &536.00 &136.53 &349.47 &0.3305 & Yes & \citet{Bhandari:2021pvj}\\
  20180916B & $29.50$ & $65.72$ &348.80 &168.73 &130.07 &0.0337 & Yes & \citet{Marcote:2020ljw}\\
  20180924B & $326.11$ & $-40.90$ &362.16 &41.45 &270.71 &0.3214 & No & \citet{Bannister:2019iju}\\
  20181030A & $158.60$ & $73.76$ &103.50 &40.16 &13.34 &0.0039 & Yes & \citet{Bhardwaj:2021hgc}\\
  20181112A & $327.35$ & $-52.97$ &589.00 &41.98 &497.02 &0.4755 & No & \citet{Prochaska:2019x}\\
  20190102C & $322.42$ & $-79.48$ &364.55 &56.22 &258.33 &0.2913 & No & \citet{Macquart:2020lln}\\
  20190523A & $207.06$ & $72.47$ &760.80 &36.74 &674.06 &0.6600 & No & \citet{Ravi:2019alc}\\
  20190608B & $334.02$ & $-7.90$ &340.05 &37.81 &252.24 &0.1178 & No & \citet{Macquart:2020lln}\\
  20190611B & $320.74$ & $-79.40$ &332.63 &56.60 &226.03 &0.3778 & No & \citet{Macquart:2020lln}\\
  20190711A & $329.42$ & $-80.36$ &592.60 &55.37 &487.23 &0.5217 & Yes & \citet{Macquart:2020lln}\\
  20190714A & $183.98$ & $-13.02$ &504.13 &38.00 &416.13 &0.2365 & No & \citet{Heintz_2020}\\
  20191001A & $323.35$ & $-54.75$ &507.90 &44.22 &413.68 &0.2340 & No & \citet{Heintz_2020}\\
  20191228A & $344.43$ & $-29.59$ &297.50 &33.75 &213.75 &0.2432 & No &\citet{Bhandari:2021pvj}\\
  20200430A & $229.71$ & $12.38$ &380.25 &27.35 &302.90 &0.1608 & No & \citet{Bhandari:2021pvj}\\
  20200906A & $53.50$ & $-14.08$ &577.80 &36.19 &491.61 &0.3688 & No & \citet{Bhandari:2021pvj}\\
  20201124A & $77.01$ & $26.06$ &413.52 &126.49 &237.03 &0.0979 & Yes &\citet{Fong:2021xxj}\\
  \hline
  \end{tabular}}
\end{table}

\begin{figure}
  \centering
  \includegraphics[width=0.32\textwidth]{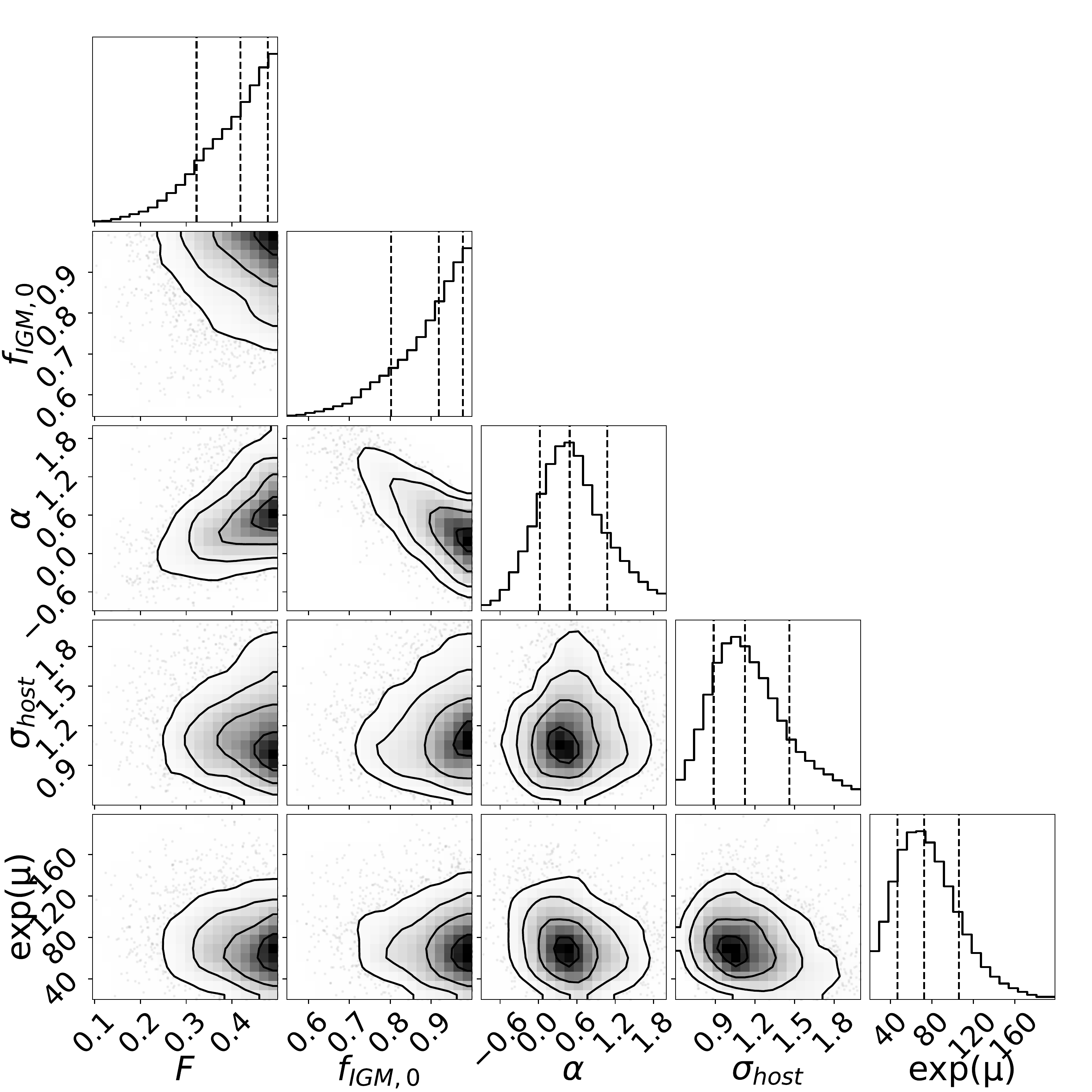}
  \includegraphics[width=0.32\textwidth]{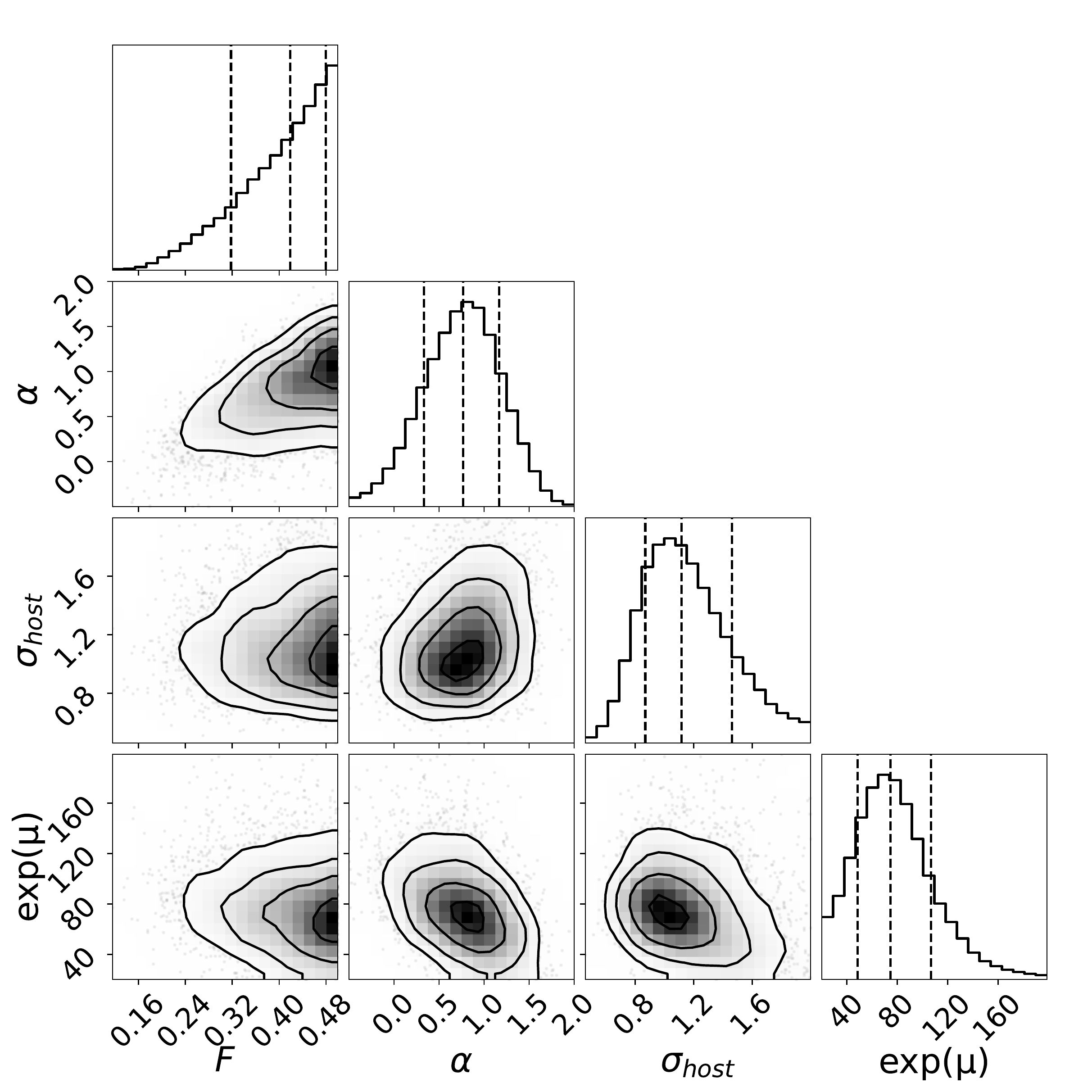}
  \includegraphics[width=0.32\textwidth]{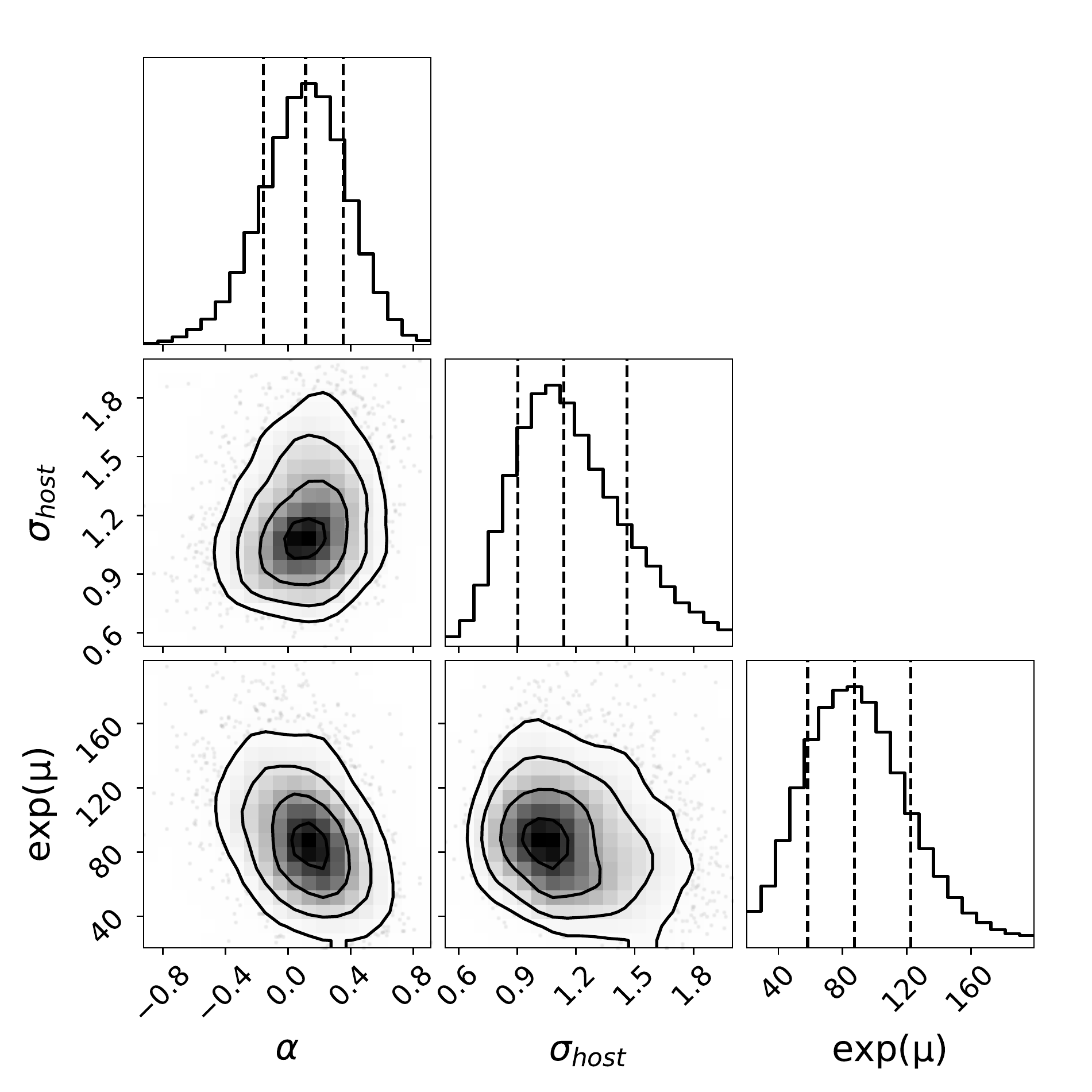}
  \caption{The contour plots constrained from 18 well-localized FRBs. The dashed lines from left to right in each subfigure represent the $16\%$, $50\%$ and $84\%$ quantiles of the distribution, respectively. Left panel: constraints on five parameters; Middle panel: constraints on four parameters with $f_\mathrm{IGM,0} = 0.84$ fixed; Right panel: constraints on three parameters with $f_\mathrm{IGM,0} = 0.84$ and $F = 0.2 $ fixed.}\label{18samples_plot}
\end{figure}

In the ideal case, we hope that all the five parameters ($F$, $f_\mathrm{IGM,0}$, $ \alpha$, $\sigma_{\rm host}$, $\mathrm{exp}(\mu)$) can be simultaneously constrained well. In practice, we use $\mathrm{exp}(\mu)$ instead of $\mu$ as a free parameter, because $\mathrm{exp}(\mu)$ directly represents the median value of $\mathrm{DM_{host}}$. The posterior probability density functions of the free parameters are calculated with the publicly available Python package code \textsf{emcee} \citep{Foreman-Mackey:2012any}. Flat priors are used for all the five parameters: $F \in U(0.01,0.5)$, $f_\mathrm{IGM,0} \in U(0,1)$, $\alpha \in U(-2,2)$, $\sigma_\mathrm{host} \in U(0.2,2)$, and $\mathrm{exp}(\mu) \in U(20,200)~\mathrm{pc ~ cm ^ {-3}}$. The 2D marginalized posterior distributions and the $1-3\sigma$ confidence contours of the five parameters are plotted in the left panel of Figure \ref{18samples_plot}, together with the best-fitting parameters listed in Table \ref{18samples_table}. One can see that although $\sigma_\mathrm{host}$ and $\mathrm{exp}(\mu)$ are tightly constrained, the constraints on $F$, $f_\mathrm{IGM,0}$ and $\alpha$ are not strict. One possible factor is that the FRB sample is not large enough to simultaneously constrain a model with too many free parameters. To reduce the freedom, we try to fix one or two parameters that can be constrained from other observations, to see if the FRB sample can strictly constrain the remaining parameters or not.

\begin{table}
  \centering
  \caption{The best-fitting parameters ($F$, $f_\mathrm{IGM,0}$, $ \alpha$, $\sigma_{\rm host}$, $\mathrm{exp}(\mu)$) constrained from 18 well-localized FRBs. The uncertainties are given at $1\sigma$ confidence level.}\label{18samples_table}
  {\begin{tabular}{ccccc}
    \hline\hline
    $F$ & $f_\mathrm{IGM,0}$ & $\alpha$ & $\sigma_\mathrm{host}$ & $\mathrm{exp}(\mu)/\mathrm{pc ~ cm ^ {-3}}$\\
    \hline
    $0.42_{-0.10}^{+0.06}$ & $0.92^{+0.06}_{-0.12}$ & $0.49^{+0.59}_{-0.47}$ & $1.13^{+0.33}_{-0.24}$ & $72.49^{+33.31}_{-25.62}$\\
    $0.42_{-0.10}^{+0.06}$ & $0.84$(fixed) & $0.77^{+0.40}_{-0.44}$ & $1.18^{+0.34}_{-0.25}$ & $74.63^{+32.05}_{-26.13}$\\
    $0.2$(fixed) & $0.84$(fixed) & $0.11^{+0.24}_{-0.27}$ & $1.14^{+0.32}_{-0.23}$ & $87.44^{+34.86}_{-29.16}$\\
    \hline
  \end{tabular}}
\end{table}

The fraction of baryon mass in IGM in the local Universe, $ f_\mathrm{IGM,0}$, has been constrained by various observations. For instance, \citet{Fukugita:1997bi} obtained $f_\mathrm{IGM,0} \approx 0.83$ from directly observing the budget of baryons in different states. \citet{Ferrara:2014sda} investigated the reionization history of IGM, and estimated that $f_\mathrm{IGM}\approx 0.82$ at $z\lesssim 0.4$, which mildly increases to $f_\mathrm{IGM}\approx0.9$ at $z\gtrsim 1.5$. Based on five well-localized FRBs, \citet{Li:2020qei} obtained $ f_\mathrm{IGM,0} = 0.84 ^{+0.16}_{-0.22}$, but no strong evidence for redshift evolution was found. All these results are consistent with each other within uncertainty. Therefore, to reduce the freedom, we first fix $f_\mathrm{IGM,0} = 0.84$ and constrain the remaining four parameters. The confidence contours and the marginalized probability distributions of the remaining parameters are plotted in the middle panel of Figure \ref{18samples_plot}, and the best-fitting parameters are summarized in Table \ref{18samples_table}. We see that the uncertainty on $\alpha$ is slightly reduced, but the constraints on the other parameters hardly change. Especially, the parameter $F$ is still not well constrained.

Actually, the parameter $F$ prefers a value around $F = 0.2$ or even smaller at $z\leq 1$ \citep{Jaroszynski:2018vgh,Kumar:2019qhc}. However, the constraint from our FRB sample prefers a much larger value, $F\gtrsim 0.4$, which may cause bias on the other parameters. Therefore, we additionally fix $F = 0.2$, and keep the other three parameters ($\alpha$, $\sigma_{\rm host}$, $\mathrm{exp}(\mu)$) free. The contour plots are demonstrated in the right panel of Figure \ref{18samples_plot}, and the best-fitting parameters are summarized in Table \ref{18samples_table}. As can be seen, fixing $F$ further reduces the uncertainty on $\alpha$, but does not strongly affect the uncertainties on $\sigma_{\rm host}$ and  $\mathrm{exp}(\mu)$. In addition, fixing $F$ strongly affects the central values of $\alpha$ and $\mathrm{exp}(\mu)$, but does not significantly affect the central value of $\sigma_{\rm host}$. The result of $\alpha = 0.11^{+0.24}_{-0.27}$ is consistent with the truth that $f_\mathrm{IGM}$ may slowly increase with redshift. But due to the large uncertainty, it is still consistent with no redshift evolution. A much larger FRB sample in a wider redshift range may help us to further reduce the uncertainty. To check this, we perform Monte Carlo simulations, which will be discussed in detail in the next section.

\section{MONTE CARLO SIMULATIONS}\label{sec_3}

With the progress of detection technique, more and more FRBs are expected to be discovered in the future, and a fraction of them can be well-localized. In addition, FRBs at high redshift may be detectable as the improvement of sensitivity of radio telescopes such as the FAST \citep{Nan:2011um}. Therefore, it is interesting to investigate the constraining ability on cosmological parameters if a large sample of FRBs in a wide redshift range are available. To this end, we perform Monte Carlo simulations to test the efficiency of our method.

The intrinsic redshift distribution of FRBs is still unclear due to the small well-localized sample. Several possibilities have been discussed in some papers. For example, \citet{Yu:2017beg} assumed that FRBs may have a similar redshift distribution to gamma-ray bursts, \citet{Li:2019klc} assumed that FRBs have a constant comoving number density but with a Gaussian cutoff, \citet{Zhang:2020ass} argued that the redshift distribution of FRBs is expected to be related with SFR, or influenced by the compact star merger but with an additional time delay. In this paper, we adopt the SFR-related model, in which the probability density function takes the form \citep{Zhang:2020ass}

\begin{equation}\label{p_z}
  P(z)  \propto \frac{4\pi D_c^2(z)\mathrm{SFR}(z)}{(1+z)H(z)},
\end{equation}
where $D_c(z) = \int_{0}^{z}\frac{cdz}{H(z)} $ represents the comoving distance, with $c$ the speed of light and $H(z) = H_0\sqrt{\Omega_m(1+z)^3+\Omega_\Lambda}$ the Hubble expansion rate, and the SFR takes the form \citep{Yuksel:2008cu}
\begin{equation}
  \mathrm{SFR}(z) = 0.02\left[(1+z)^{a\eta} + \left(\frac{1+z}{B}\right)^{b\eta} + \left(\frac{1+z}{C}\right)^{c\eta}\right]^{1/\eta},
\end{equation}
where $a = 3.4$, $b = -0.3$, $c = -3.5$, $B = 5000$, $C = 9$ and $\eta = -10$.

We simulate a set of mock FRB samples, and use them to constrain the parameters ($F$, $\alpha$, $\sigma_\mathrm{host}$,$\mathrm{exp}(\mu)$). The simulations are performed based on the standard $\mathrm{\Lambda CDM}$ model with Planck 2018 parameters \citep{Planck:2018vyg}. The other fiducial parameters are $F = 0.2$, $f_\mathrm{IGM,0} = 0.84$, $\alpha = 0 ~ \mathrm{or} ~ 0.2$, $\sigma_\mathrm{host} = 1.0$ and $\mathrm{exp}(\mu) = 100~\mathrm{pc ~ cm ^ {-3}}$. The main procedures of the simulations are described below.
\begin{enumerate}[(1)]
\item Randomly draw a certain number of redshifts from equation (\ref{p_z}). We set $z_{\rm max}  = 3$.
\item Calculate $\langle \mathrm{DM_{IGM}}(z)\rangle $ according to equation (\ref{DM_IGM_average}).
\item Randomly draw the same number of $\Delta$ from equation (\ref{p_delta}).
\item Calculate $\mathrm{DM_{IGM}}=\Delta\times\langle \mathrm{DM_{IGM}}(z)\rangle $.
\item Randomly draw the same number of $\mathrm{DM_{host}}$ from equation (\ref{p_dm_host})
\item Calculate $\mathrm{DM_E}$ according to the second equality of equation (\ref{DM_E}).
\item Finally, we obtain a sample of mock FRBs ($z_i,\mathrm{DM_E}_i$), $i=1,2,\cdots, N$.
\end{enumerate}

\begin{figure}
  \centering
  \includegraphics[width=0.32\textwidth]{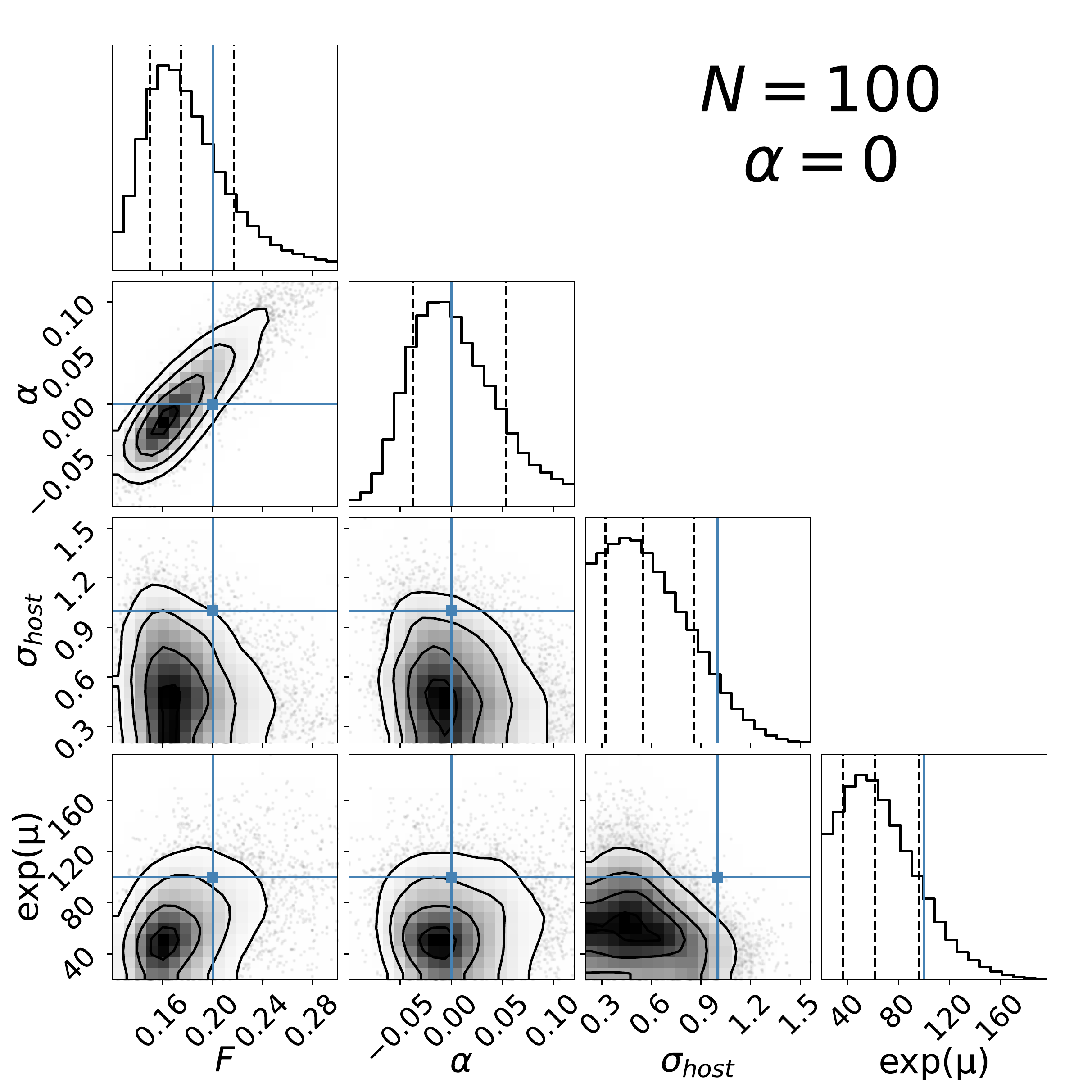}
  \includegraphics[width=0.32\textwidth]{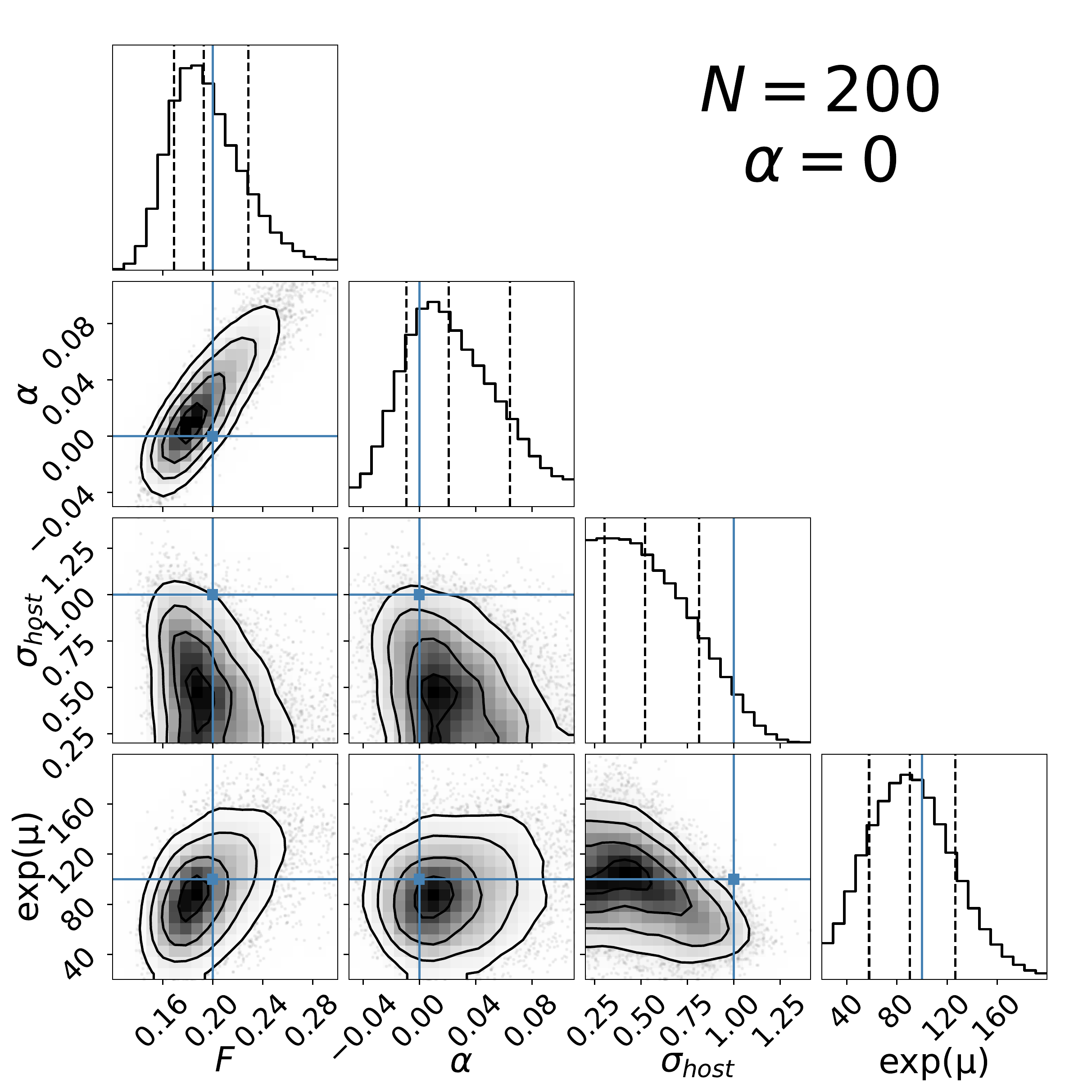}
  \includegraphics[width=0.32\textwidth]{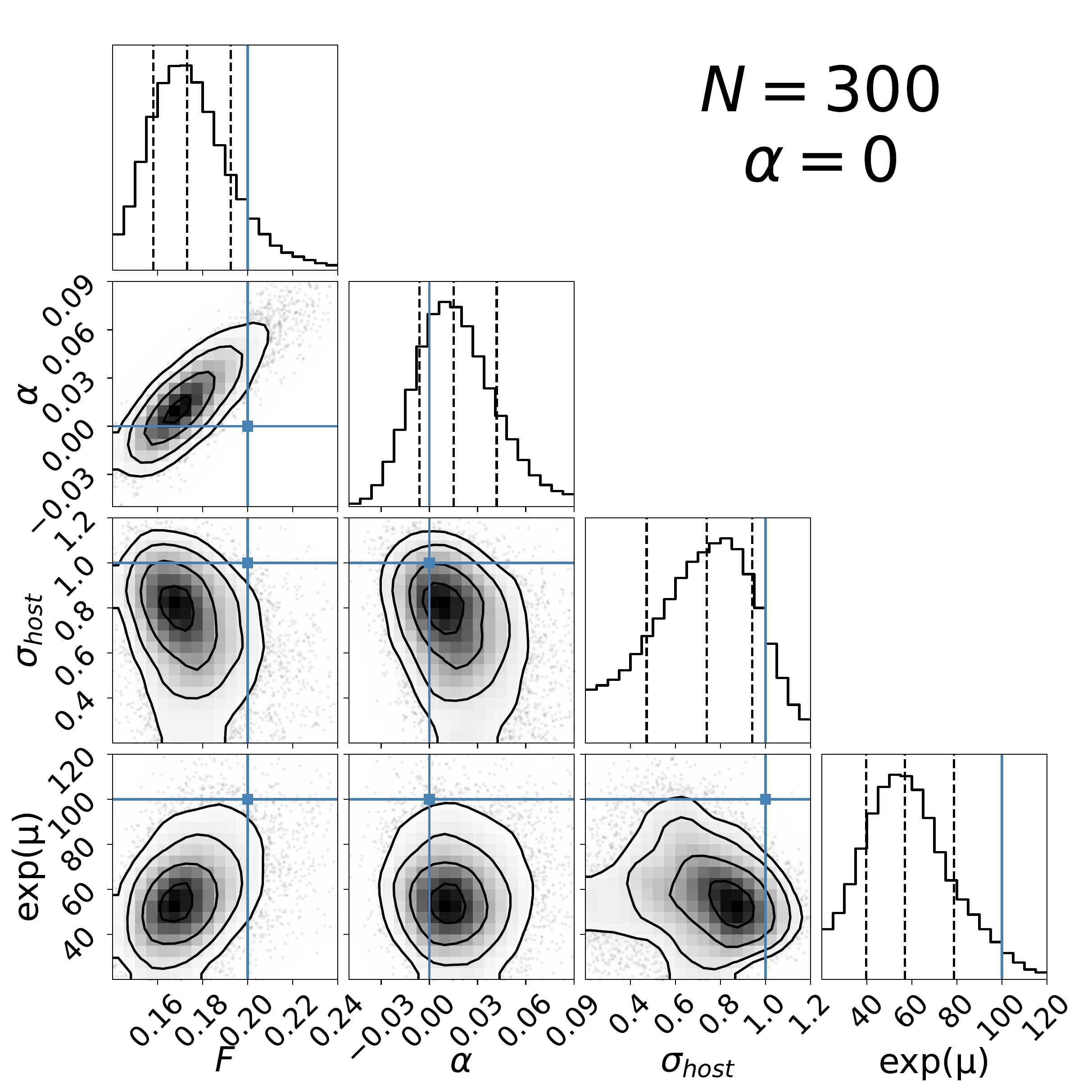}
  \includegraphics[width=0.32\textwidth]{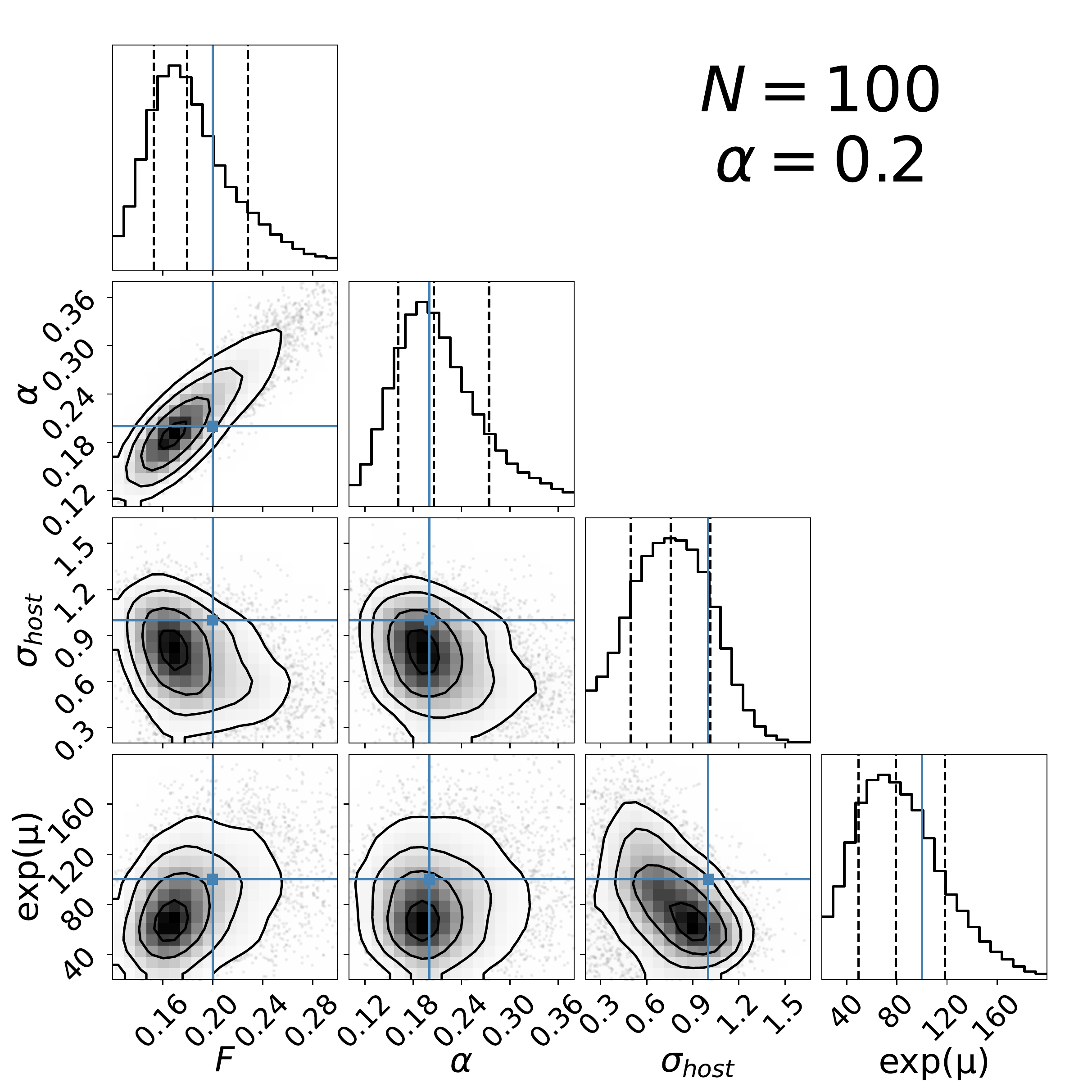}
  \includegraphics[width=0.32\textwidth]{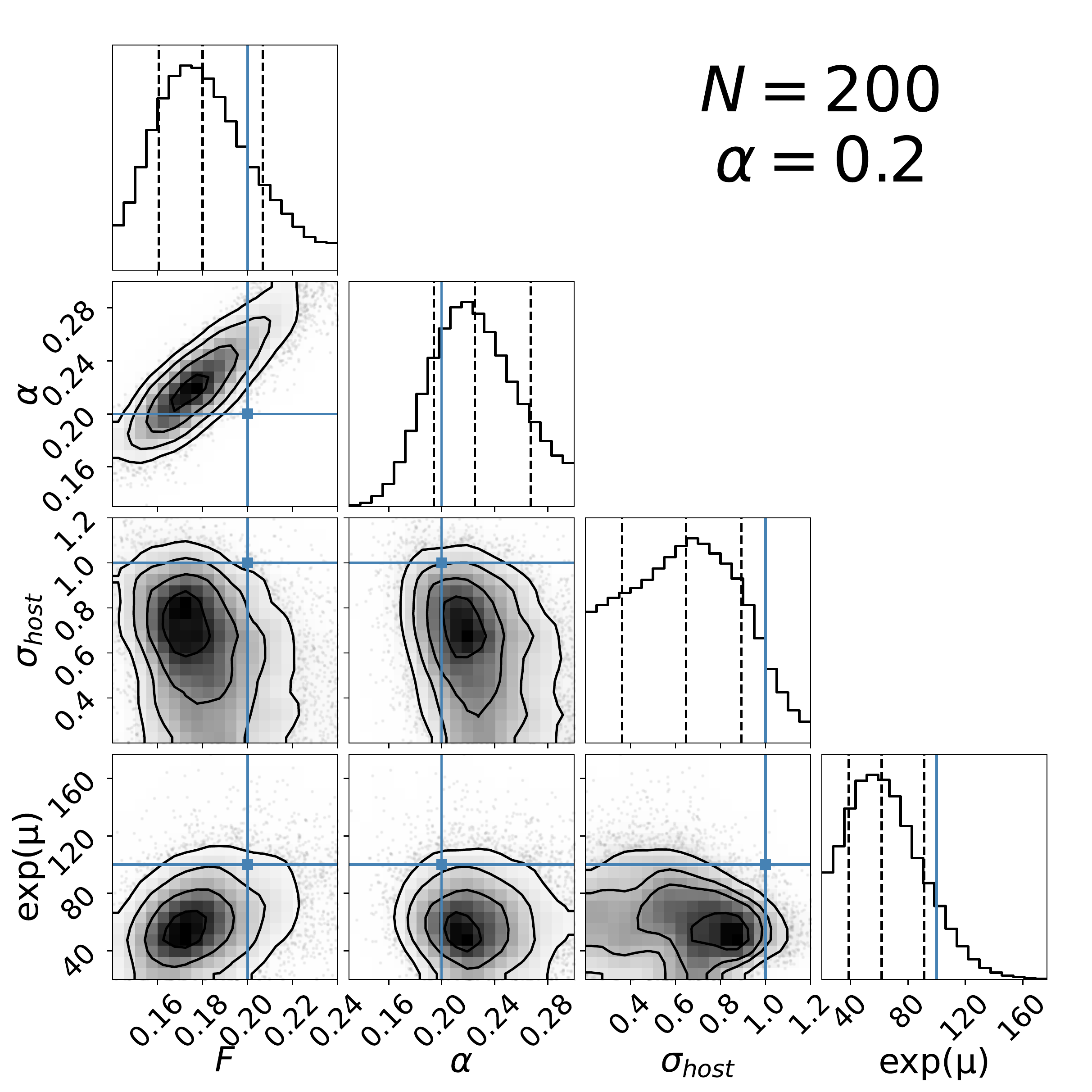}
  \includegraphics[width=0.32\textwidth]{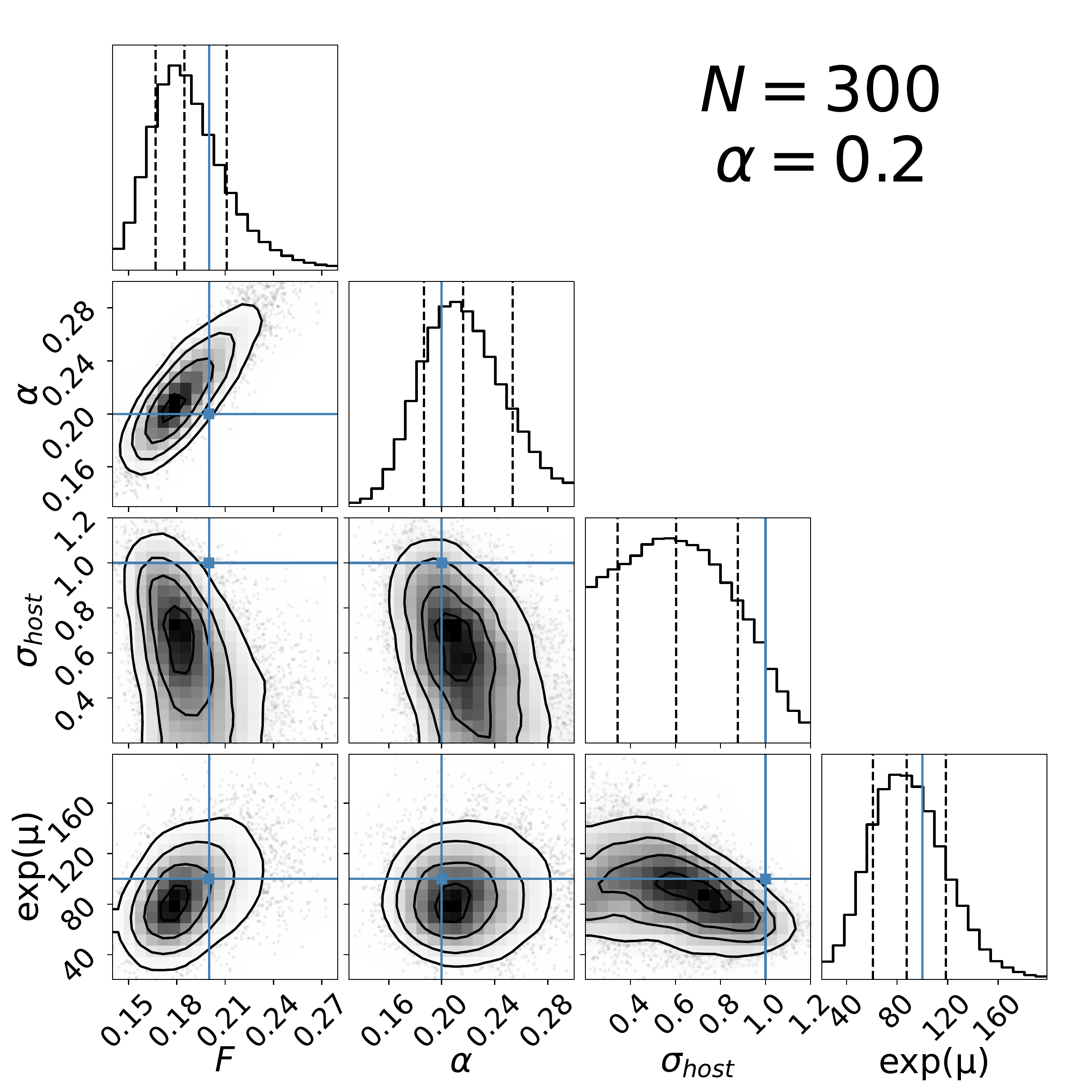}
  \caption{Constraints on four parameters ($F$, $\alpha$, $\sigma_\mathrm{host}$, $\mathrm{exp}(\mu)$) by using $N=100$, 200, 300 (from left to right) mock FRBs, respectively. The fiducial values are $F = 0.2$, $\alpha = 0$ (top) and $\alpha = 0.2$ (bottom), $\sigma_\mathrm{host} = 1.0$ and $\mathrm{exp}(\mu) = 100~\mathrm{pc ~ cm ^ {-3}}$, which are reflected by the solid, blue lines. The dashed lines from left to right represent the $16\%$, $50\%$ and $84\%$ quantiles of the distribution, respectively.}\label{simulated_4p_plot}
\end{figure}

We replace the true FRB samples with the mock samples, and use them to constrain the parameters as described above. First, we use the mock FRB samples to test the efficiency of our four-parameter model (with $f_\mathrm{IGM,0} = 0.84$ fixed). The corresponding contour plots with $N=100$, 200 and 300 FRBs are shown in Figure \ref{simulated_4p_plot}. We consider two different fiducial values of $\alpha$, i.e. $\alpha = 0 $ (top three panels) and $\alpha = 0.2$ (bottom three panels). We summarize the best-fitting parameters and their $1\sigma$ uncertainties in Table \ref{simulated_4p_table}. With the mock FRB samples, all the four parameters can be well constrained. However, we note that although $F$ and $\alpha$ correctly recover their fiducial values within $1\sigma$ uncertainty in most cases, $\sigma_{\rm host}$ and $\mathrm{exp}(\mu)$ are usually biased. The best-fitting values of $\sigma_{\rm host}$ and $\mathrm{exp}(\mu)$ are usually smaller than their fiducial values. Two reasons may lead to this situation. First, $\sigma_{\rm host}$ and $\mathrm{exp}(\mu)$ are two parameters related to $\mathrm{DM_{host}}$, which gets less weight in $\mathrm{DM_E}$ at higher redshift. This results in less sensitivity on $\sigma_{\rm host}$ and $\mathrm{exp}(\mu)$ in the Bayesian inference model. Second, the parameters are correlated. A biased estimation on one parameter may cause bias on the other parameters. We hope that fix one parameter (for example $F$) may alleviate the bias.

\begin{table}
  \centering
  \caption{The best-fitting parameters ($F$, $\alpha$, $\sigma_\mathrm{host}$, $\mathrm{exp}(\mu)$) constrained from $N=100$, 200 and 300 mock FRBs. The fiducial values are $F = 0.2$, $\alpha = 0$ (left panel) and  $\alpha = 0.2$ (right panel), $\sigma_\mathrm{host} = 1.0$ and $\mathrm{exp}(\mu) = 100~\mathrm{pc ~ cm ^ {-3}}$. The uncertainties are given at $1\sigma$ confidence level.}\label{simulated_4p_table}
  {\begin{tabular}{ccccc}
    \hline\hline
    $N$ & $F$ & $\alpha$ & $\sigma_\mathrm{host}$ & $\mathrm{exp}(\mu)/\mathrm{pc ~ cm ^ {-3}}$\\
    \hline
    100 & $0.17^{+0.04}_{-0.03}$ &$0.00^{+0.05}_{-0.04}$ & $0.55^{+0.31}_{-0.23}$ & $61.45^{+34.61}_{-24.95}$\\
    200 & $0.19^{+0.04}_{-0.02}$ &$0.02^{+0.04}_{-0.03}$ & $0.52^{+0.29}_{-0.22}$ & $90.25^{+36.09}_{-32.25}$\\
    300 & $0.17^{+0.02}_{-0.01}$ &$0.02^{+0.03}_{-0.02}$ & $0.74^{+0.20}_{-0.27}$ & $56.84^{+21.88}_{-17.19}$\\
    \hline
  \end{tabular}}
  {\begin{tabular}{ccccc}
    \hline\hline
    $N$ & $F$ & $\alpha$ & $\sigma_\mathrm{host}$ & $\mathrm{exp}(\mu)/\mathrm{pc ~ cm ^ {-3}}$\\
    \hline
    100 & $0.18^{+0.05}_{-0.03}$ &$0.21^{+0.07}_{-0.04}$ & $0.76^{+0.26}_{-0.26}$ & $79.21^{+39.03}_{-29.77}$\\
    200 & $0.18^{+0.03}_{-0.02}$ &$0.23^{+0.04}_{-0.03}$ & $0.65^{+0.25}_{-0.28}$ & $61.76^{+29.61}_{-22.99}$\\
    300 & $0.19^{+0.03}_{-0.02}$ &$0.22^{+0.04}_{-0.03}$ & $0.60^{+0.28}_{-0.26}$ & $87.58^{+30.96}_{-26.59}$\\
    \hline
  \end{tabular}}
\end{table}

\begin{figure}
  \centering
  \includegraphics[width=0.32\textwidth]{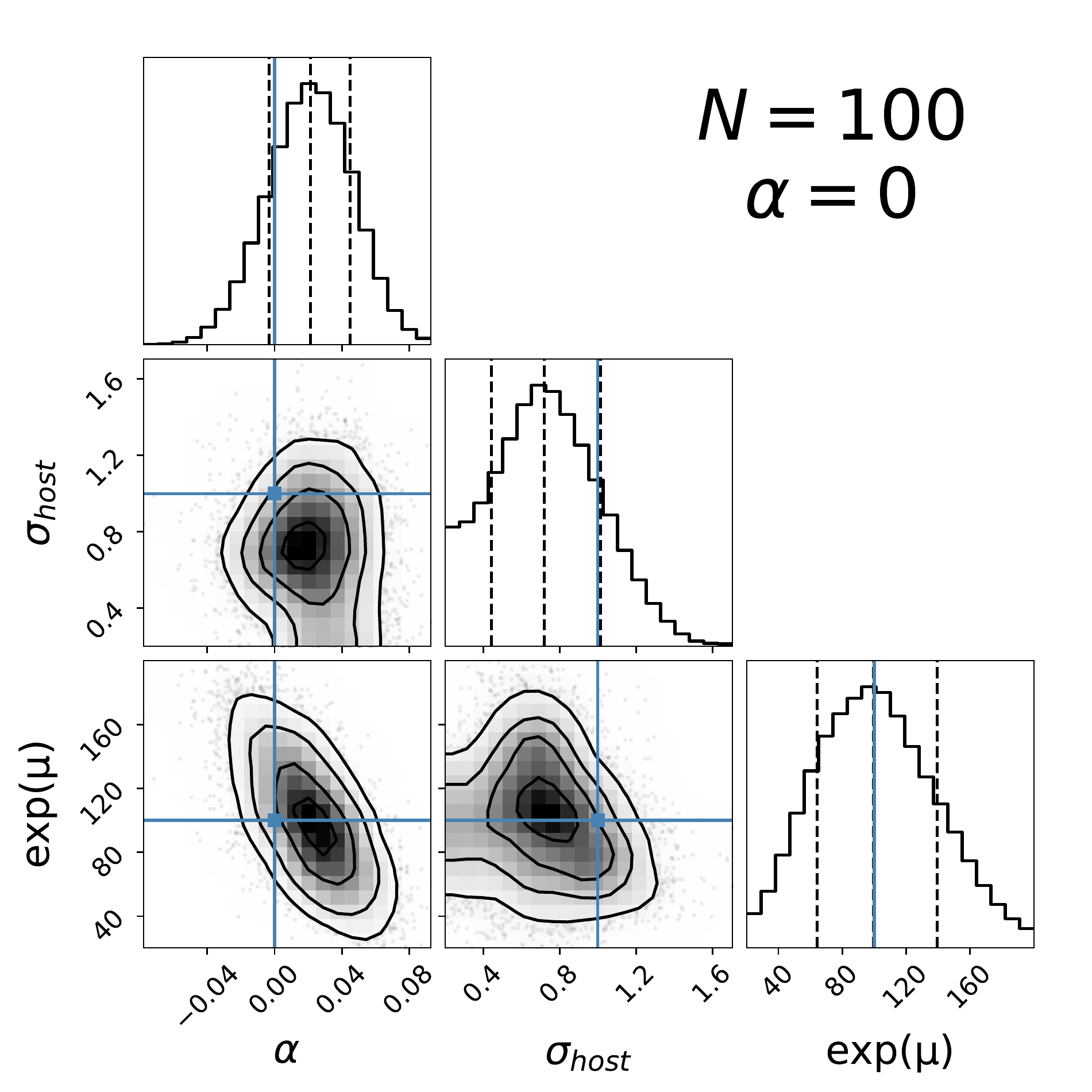}
  \includegraphics[width=0.32\textwidth]{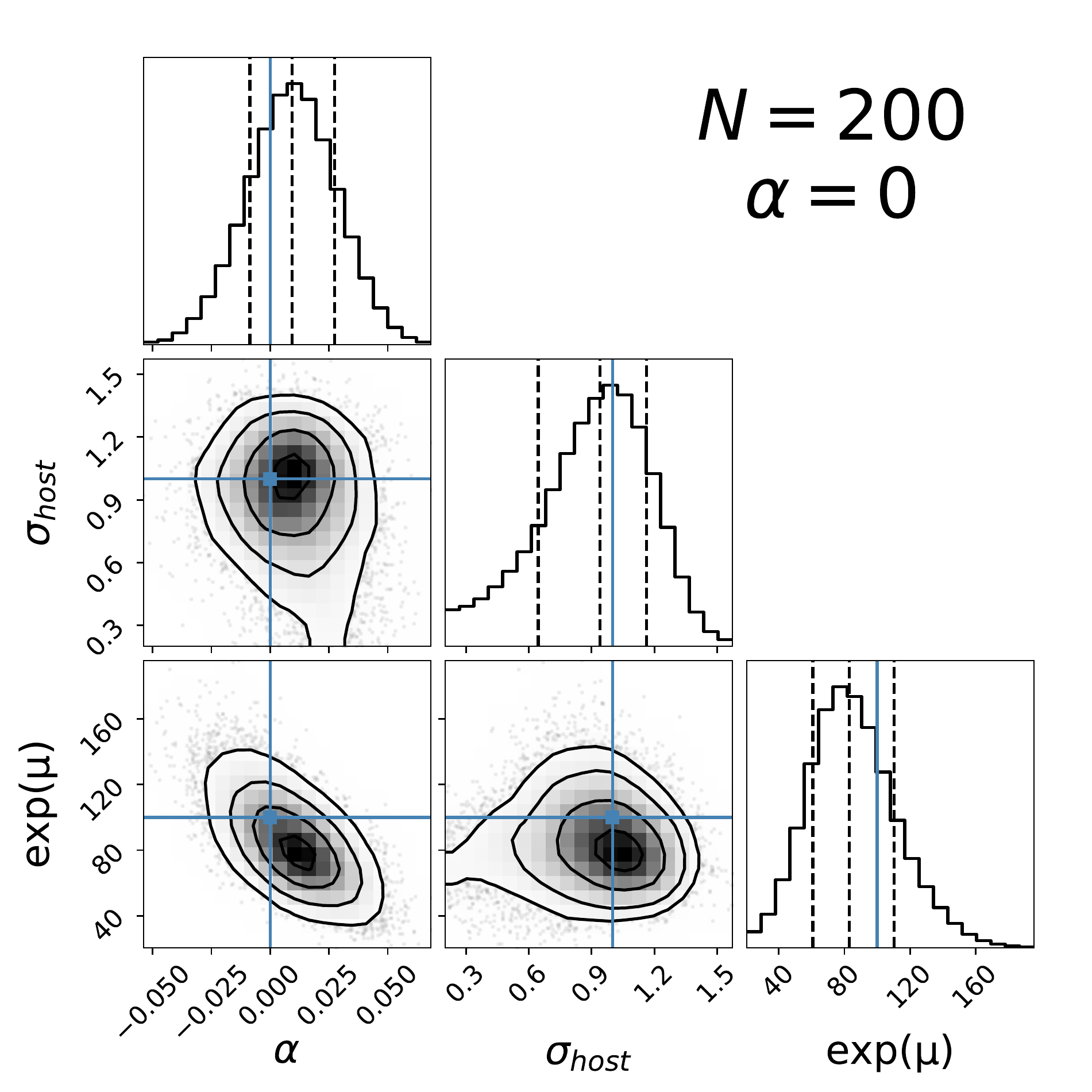}
  \includegraphics[width=0.32\textwidth]{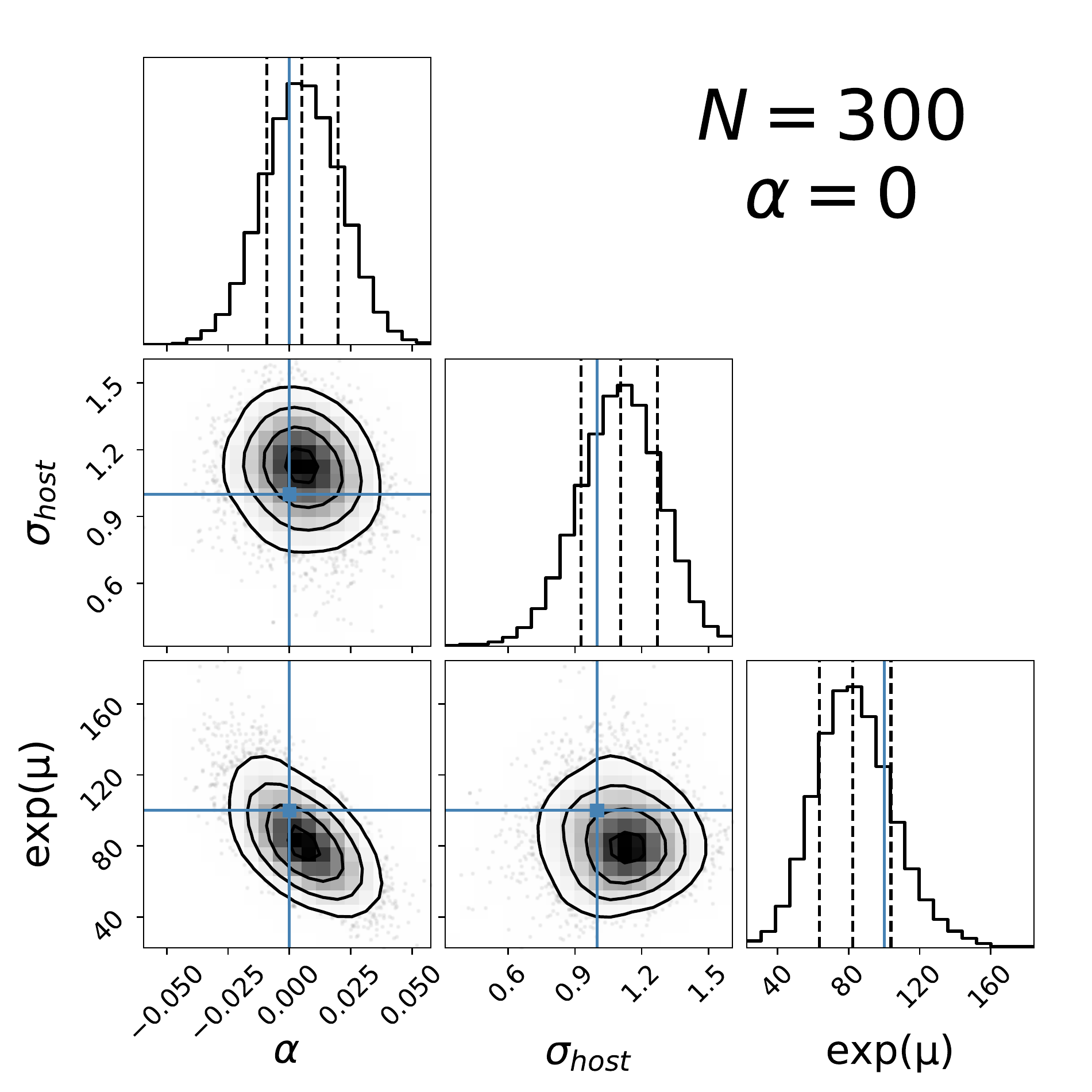}
  \includegraphics[width=0.32\textwidth]{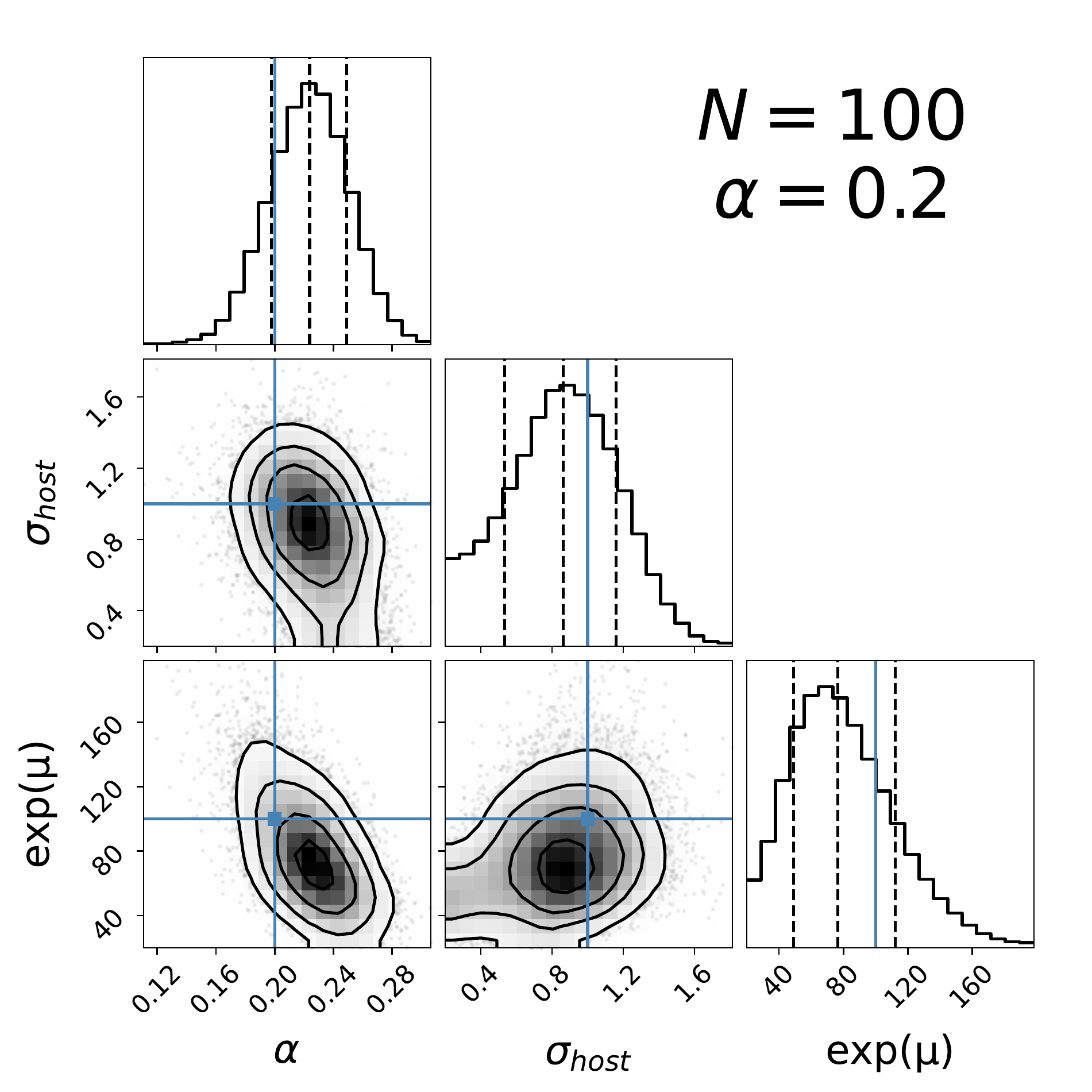}
  \includegraphics[width=0.32\textwidth]{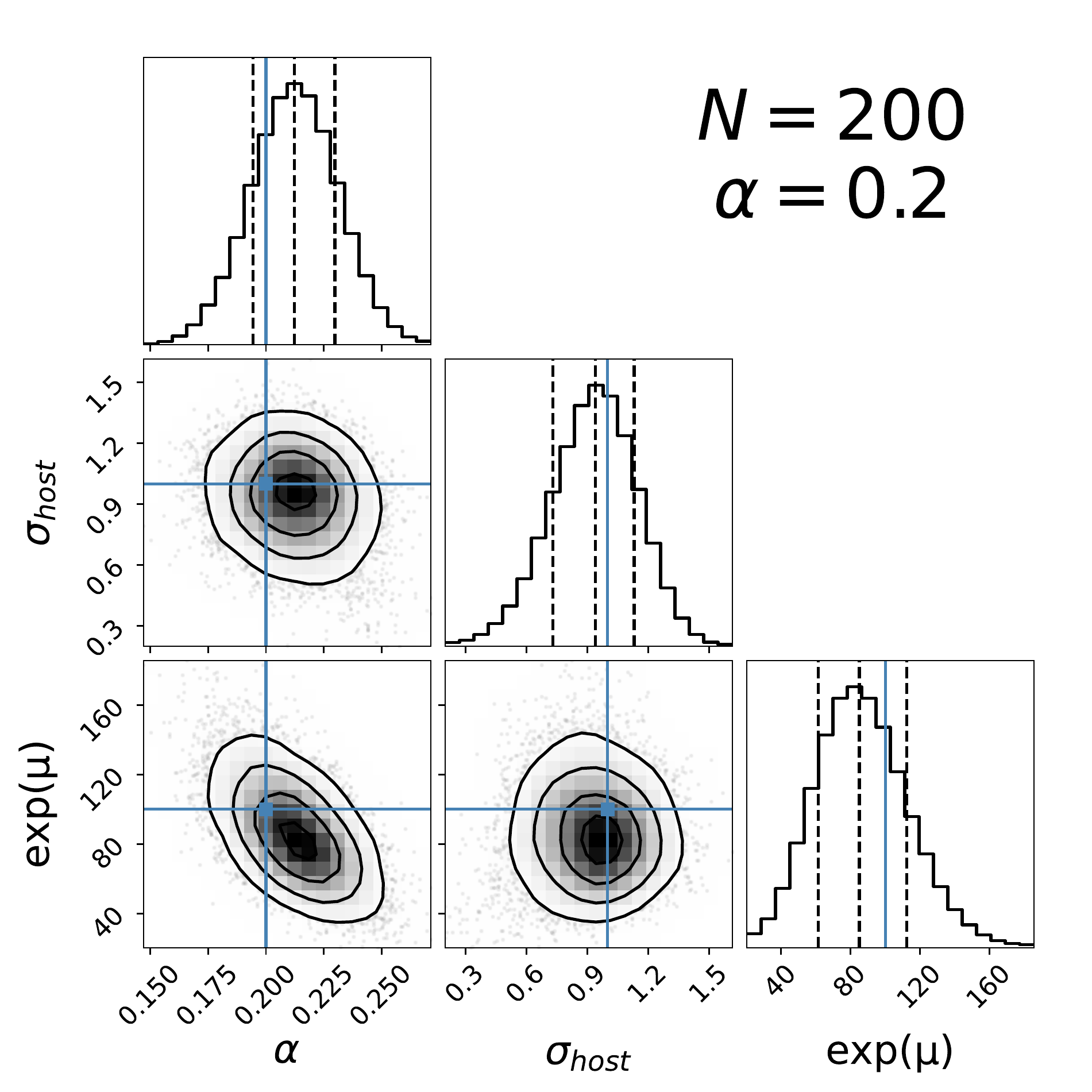}
  \includegraphics[width=0.32\textwidth]{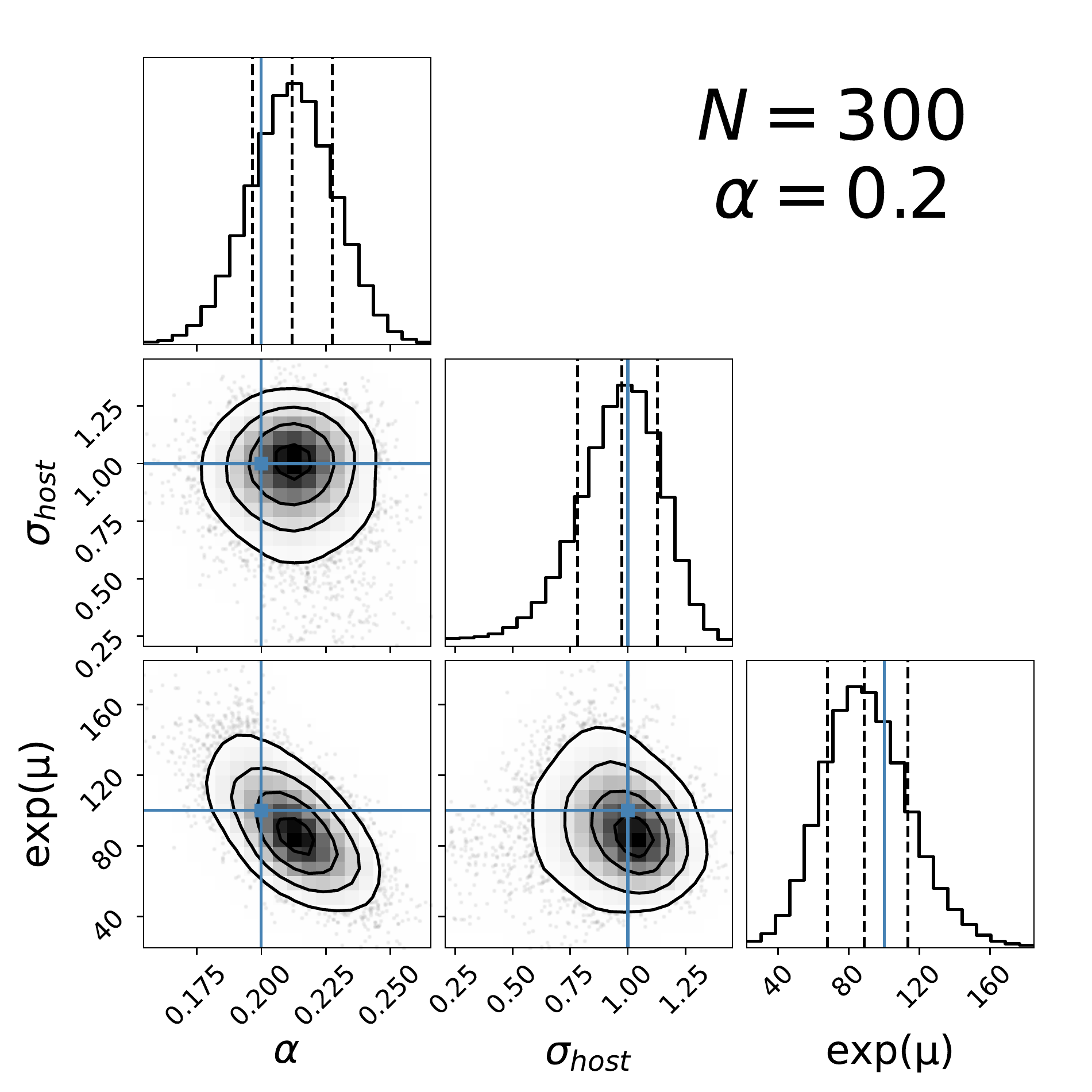}
  \caption{Constraints on three parameters ($\alpha$, $\sigma_\mathrm{host}$, $\mathrm{exp}(\mu)$) by using $N=100$, 200, 300 (from left to right) mock FRBs, respectively. The fiducial values are $F = 0.2$, $\alpha = 0$ (top) and  $\alpha = 0.2$ (bottom), $\sigma_\mathrm{host} = 1.0$ and $\mathrm{exp}(\mu) = 100~\mathrm{pc ~ cm ^ {-3}}$, which are reflected by the solid, blue lines. The dashed lines from left to right represent the $16\%$, $50\%$ and $84\%$ quantiles of the distribution, respectively.}\label{simulated_3p_plot}
\end{figure}

To check this, we additionally fix $F=0.2$, and use the same mock samples to constrain three parameters ($\alpha$, $\sigma_\mathrm{host}$, $\mathrm{exp}(\mu)$) in the identical manner to the previous four-parameter model. Through the combined contour plots (Figure \ref{simulated_3p_plot}) and the best-fitting results (Table \ref{simulated_3p_table}), one can see that all the parameters correctly recover the fiducial values within $1\sigma$ uncertainty. Compared with the true FRB samples, the parameter $\alpha$ is more tightly constrained, thanks to the enlargement of FRB sample and the extension of redshift range. In addition, the uncertainties on $\sigma_\mathrm{host}$ and $\mathrm{exp}(\mu)$ are reduced gradually with the increase of FRB number.

\begin{table}
  \centering
  \caption{The best-fitting parameters ($\alpha$, $\sigma_\mathrm{host}$, $\mathrm{exp}(\mu)$) constrained from $N=100$, 200 and 300 mock FRBs. The fiducial values are $F = 0.2$, $\alpha = 0$ (left panel) and  $\alpha = 0.2$ (right panel), $\sigma_\mathrm{host} = 1.0$ and $\mathrm{exp}(\mu) = 100~\mathrm{pc ~ cm ^ {-3}}$. The uncertainties are given at $1\sigma$ level.}\label{simulated_3p_table}
  {\begin{tabular}{cccc}
    \hline\hline
    $N$ & $\alpha$ & $\sigma_\mathrm{host}$ & $\mathrm{exp}(\mu)/\mathrm{pc ~ cm ^ {-3}}$\\
    \hline
    100 & $0.02^{+0.02}_{-0.02}$ & $0.72^{+0.29}_{-0.28}$ & $99.49^{+39.89}_{-35.38}$\\
    200 & $0.01^{+0.02}_{-0.02}$ & $0.94^{+0.22}_{-0.29}$ & $82.90^{+27.14}_{-21.99}$\\
    300 & $0.01^{+0.01}_{-0.01}$ & $1.11^{+0.17}_{-0.18}$ & $82.24^{+21.55}_{-18.77}$\\
    \hline
  \end{tabular}}
  {\begin{tabular}{cccc}
    \hline\hline
    $N$ & $\alpha$ & $\sigma_\mathrm{host}$ & $\mathrm{exp}(\mu)/\mathrm{pc ~ cm ^ {-3}}$\\
    \hline
    100 & $0.22^{+0.03}_{-0.03}$ & $0.86^{+0.30}_{-0.34}$ & $76.87^{+35.41}_{-27.48}$\\
    200 & $0.21^{+0.02}_{-0.02}$ & $0.94^{+0.19}_{-0.21}$ & $85.07^{+27.11}_{-23.51}$\\
    300 & $0.21^{+0.02}_{-0.02}$ & $0.97^{+0.15}_{-0.19}$ & $88.76^{+24.46}_{-20.78}$\\
    \hline
  \end{tabular}}
\end{table}

\section{DISCUSSION AND CONCLUSIONS}\label{sec_4}

In this paper, we investigate the baryon mass fraction in IGM using well-localized FRBs. Considering the probability distributions of $\mathrm{DM_{IGM}}$ and $\mathrm{DM_{host}}$, we construct a five-parameter Bayesian inference model. The free parameters ($F$, $f_\mathrm{IGM,0}$, $\alpha$, $\sigma_\mathrm{host}$, $\mathrm{exp}(\mu)$) are constrained using 18 well-localized FRBs that have direct redshift measurement. Unfortunately, due to the small FRB sample, the five parameters can't be well constrained simultaneously. Especially, the best-fitting value of $f_{\rm IGM,0}$ is somewhat larger than expected, and the constraint on $\alpha$ is loose. Considering that the fraction of baryon mass in the local universe can be constrained from other independent observations, we fix the parameter $f_\mathrm{IGM,0}=0.84$ and free the remaining four parameters. The uncertainty on $\alpha$ is slightly reduced in four-parameter fit, but the parameter $F$ is still not well-constrained. If we further fix $F=0.2$, the remaining three parameters can be tightly constrained, with the best-fitting results $\alpha =0.11^{+0.24}_{-0.27}$, $\sigma_\mathrm{host} = 1.14^{+0.32}_{-0.23}$ and $\mathrm{exp}(\mu) = 87.44^{+34.86}_{-29.16}~\mathrm{pc ~ cm ^ {-3}}$. The positive central value of $\alpha$ is consistent with the possibility that $f_\mathrm{IGM}$ may slowly increase with redshift, but due to the large uncertainty it is still consistent with no redshift evolution. Therefore, the present FRB sample is not large enough to prove or falsify the redshift dependence of baryon mass fraction in IGM.

With the operation of ongoing and upcoming radio telescopes, the available FRB sample is expected to be significantly enlarged in the near future. At the same time, the detectable redshift range is expected to be highly extended up to $z\approx 3$. To this end, we perform Monte Carlo simulations to check the efficiency of our method. It is found that even if we enlarge the FRB sample to $N=300$, the five parameters can't be tightly constrained simultaneously. Fixing $f_{\rm IGM,0}$ helps to improve the constraints on $\alpha$, but the parameters $F$ still can't be well constrained. In addition, the parameters $\sigma_\mathrm{host}$ and $\mathrm{exp}(\mu)$ may be biased. Only if we simultaneously fix the two parameters $f_{\rm IGM,0}$ and $F$, we can achieve an unbiased estimation on the remaining parameters. In this case, the parameter $\alpha$ are tightly constrained, at the level of $\sim 0.02$ with $N=100$. Therefore, in order to test the redshift dependence of baryon mass fraction in IGM using FRBs, precise constraints on $f_{\rm IGM,0}$ and $F$ using other independent observations are necessary.

We note that there is correlation between parameters. For instance, in the case of fixed $F$ and $f_{\rm IGM,0}$, a larger $\alpha$ value means a larger contribution of ${\rm DM_{IGM}}$ thus a smaller contribution of ${\rm DM_{host}}$, hence a smaller value of $\exp(\mu)$. Therefore, $\alpha$ and $\exp(\mu)$ are negatively correlated. This can be seen clearly from the contour plot in the right panel of Figure \ref{18samples_plot} (and also Figure \ref{simulated_3p_plot}). For the fixed $f_{\rm IGM,0}$, a larger value of $F$ means a larger variance of ${\rm DM_{IGM}}$, hence the probability of ${\rm DM_{IGM}}$ being large is higher, which leads to a larger value of $\alpha$. Therefore, $\alpha$ and $F$ are positively correlated, see the contour plot in the middle panel of Figure \ref{18samples_plot} (this can be seen more clearly from Figure \ref{simulated_4p_plot}). From the left panel of Figure \ref{18samples_plot}, we can also see that $f_{\rm IGM,0}$ is negatively correlated with $\alpha$. This is a natural result, since for a fixed ${\rm DM_{IGM}}$ value, a larger $f_{\rm IGM,0}$ value requires a smaller $\alpha$ value. The parameter correlation explains why the five parameters can't be tightly constrained simultaneously, even if the data sample is enlarged.

\section*{Acknowledgements}
This work has been supported by the National Natural Science Fund of China (Grant Nos. 12275034, 11873001 and 12147102), and the Fundamental Research Funds for the Central Universities of China (Grants No. 2022CDJXY-002).

\section*{Data Availability}
The Host/FRB catalog is available at the FRB Host Database \textcolor{blue}{http://frbhosts.org}.

\bibliographystyle{mnras}
\bibliography{reference}

\label{lastpage}
\end{document}